\documentclass[10pt,conference]{IEEEtran}
\usepackage{ctable}
\usepackage{subcaption}
\usepackage{listings}
\usepackage{array}
\newcolumntype{L}[1]{>{\raggedright\let\newline\\\arraybackslash\hspace{0pt}}m{#1}}
\newcolumntype{C}[1]{>{\centering\let\newline\\\arraybackslash\hspace{0pt}}m{#1}}
\newcolumntype{R}[1]{>{\raggedleft\let\newline\\\arraybackslash\hspace{0pt}}m{#1}}

\newcommand{\tool}{Minion} 
\newcommand{\code}[1]{{\footnotesize\textsf{#1}}}

\usepackage{listings}
\usepackage{color}

\definecolor{dkgreen}{rgb}{0,0.6,0}
\definecolor{gray}{rgb}{0.5,0.5,0.5}
\definecolor{mauve}{rgb}{0.58,0,0.82}

\begin{document}

\title{Toward Mining Visual Log of Software}


\author{\IEEEauthorblockN{Hung Viet Pham, Tam The Nguyen, Phong Minh Vu, Tung Thanh Nguyen}
\IEEEauthorblockA{Computer Science Department\\
Utah State University\\
\lbrack hung.pham, tam.nguyen, phong.vu\rbrack@aggiemail.usu.edu, tung.nguyen@usu.edu\\
}}

\maketitle

\begin{abstract}
In this paper, we define visual log of a software system as data capturing the interactions between its users and its graphic user interface (GUI), such as screen-shots and screen recordings. We vision that mining such visual log could be useful for bug reproducing and debugging, automated GUI testing, user interface designing, question answering of common usages in software support, etc. Toward that vision, we propose a core framework for mining visual log of software. This framework focuses on detecting GUI elements and changes in visual log, removing users' private data, recognizing user interactions with GUI elements, and learning GUI usage patterns. We also performed a small study on the characteristics of GUI elements in mobile apps. The findings from this study suggested several heuristics to design techniques for recognizing GUI elements and interactions.
\end{abstract}


\section{Motivation}
Graphic user interface (GUI) is an essential part of most software applications. It allows software users to interact with the underlying software applications and to perform most important tasks in an intuitive and effective way. Therefore, ensuring and improving the effectiveness of GUI-based interactions between software users and software applications (e.g. via designing and testing) is an important task for software development teams.

Recent advanced researches in software engineering show that many software engineering problems could be solved via mining developmental and operational data of software. For example, software systems often write log files into hard drives to record important information about their executions (exceptions, critical execution points, program states, and variable values, etc.). Mining those logs is useful for several software development and maintenance tasks like debugging or performance analysis~\cite{shang_2015_log,xu_2009_log,yuan_2012_log}. 

However, those software logs are often in textual format and only record internal information of the running software systems. In this paper, we propose the concept of \emph{visual log} of software, which records interactions between users and GUI of software systems. Popular types of visual log are screen-shots (images) and screen recordings (videos). Let us elaborate how mining visual log can be helpful for software engineering via some examples.


\paragraph{Bug reporting, reproducing, and re-testing}
Software bugs could occur when a professional tester or an end-user is testing or using a software system. To inform the software team about a bug, the tester or end-user will create a bug report in which he/she often describes the symptoms and consequences of the bug, the steps to reproduce it, and sometimes attached supplemental data like a screen-shot capturing the user interface of the software system when the bug happens. This is often a time-consuming task and typically difficult for end-users who do not have good inside knowledge of that software system and of software development in general. This difficulty might lead to poorly described/written bug reports which cannot be reproduced~\cite{bettenburg_2008_bugreport,bettenburg_2008_bugreport2}.

We can enhance bug reports with visual log. For example, when a bug happens, the tester/user can record the process leading to that bug as a video clip capturing the GUI of the system and attach that video clip with the bug report. Thus, the developers assigned to verify and fix the bug will have more evidence about its appearance and the process to reproduce it.

Mining video clips capturing software applications' screen can also be helpful for bug reproducing and regression testing. We vision that a mining tool for visual logs can infer the GUI elements and interactions of the user from a given video clip capturing the bug and generate a GUI test script involving those GUI elements and interactions. Then, an automated GUI testing tool can use that test script to reproduce the corresponding bug for debugging. When the bug is fixed, the automated GUI testing tool can re-execute the test script to verify if the bug has been truly fixed or not.

\paragraph{Mining users' GUI usage patterns} When using a software application, users often perform some tasks repeatedly. For example, users of a text editor would perform opening/saving files, changing font size and face, changing page margins, etc. An experienced user often uses the same sequence of GUI interactions to perform a task, e.g. saving a file by clicking on menu item \code{File} and then click on menu sub-item \code{Save}.

We define \emph{GUI usage patterns} as sequences of GUI interactions that are performed repeatedly. It is likely that a GUI usage pattern corresponds to a frequently used task. Thus, GUI usage patterns can be useful in several ways. Software developers could analyze those  usage patterns to improve the usability of the software. For example, we can create a button as a short-cut for the usage pattern of file saving involving two interactions (click on menu item \code{File} and then \code{Save}), thus, reducing the number of GUI actions to perform that task. GUI usage patterns can be used as learning materials for beginning users. For example, a beginning user can ask a question like "How to change font size". The help system of the software application can search through its store of usage patterns, look for the pattern corresponding to the asked task, and generate a helping instruction (e.g. a textual script or even a video clip illustrating that task).

We vision that we can use a mining tool to extract sequences of GUI interactions from a large collection of visual logs, which can be collected via submissions from testers in acceptance testing, or from submissions of early access and voluntary users. Then, frequent sequence mining tools can perform on the extracted sequences of GUI interactions to recover GUI usage patterns.

\begin{figure}[t!]
	\centering
	\includegraphics[width=0.45\textwidth]{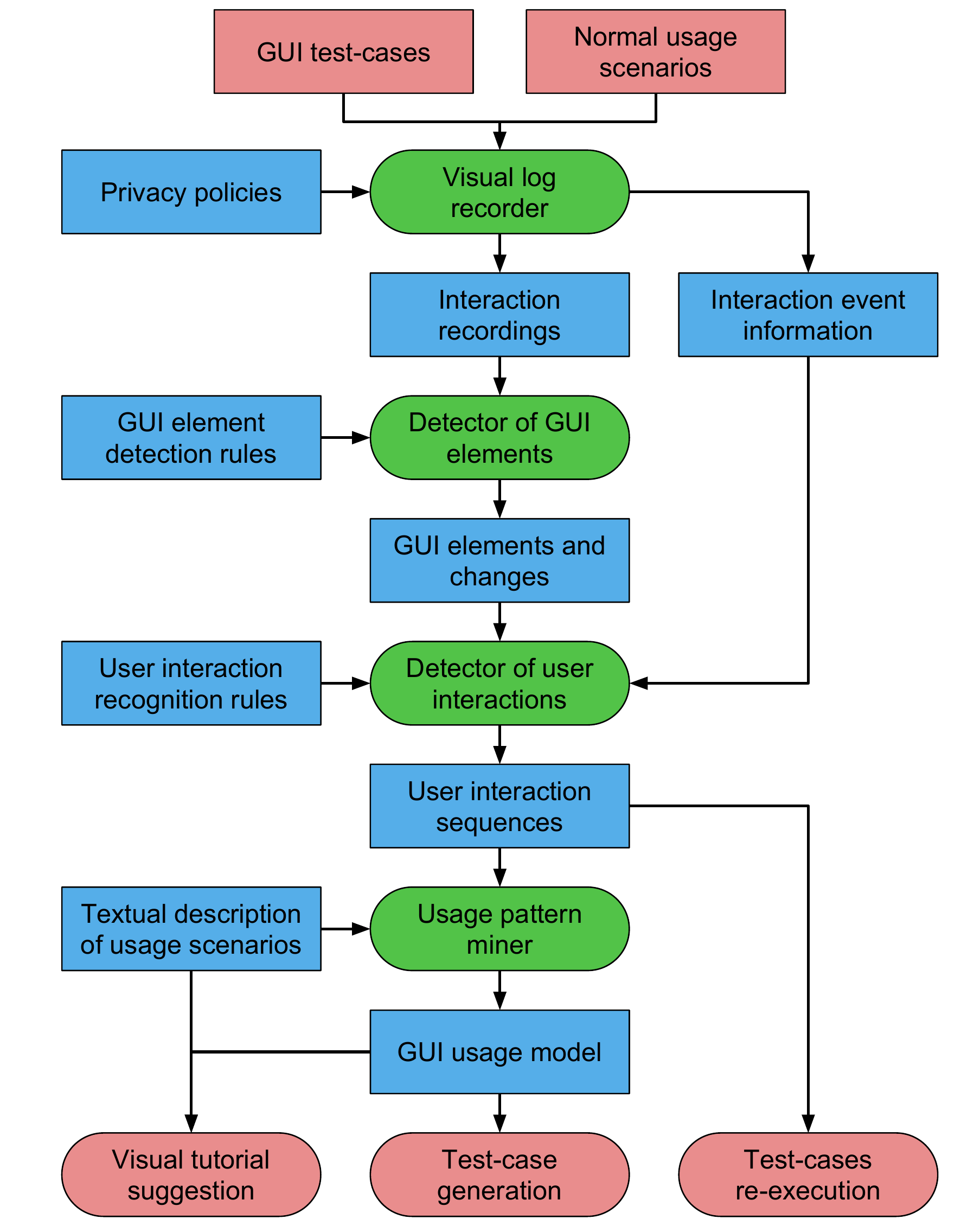}
	\caption{Architecture overview of \tool}
	\label{fig:framework}
\end{figure}

\begin{figure*}[t!]
	\centering
    \begin{subfigure}[t]{0.24\textwidth}
        \centering
        \frame{\includegraphics[width=\textwidth,clip=true,trim=0 0 5 0]{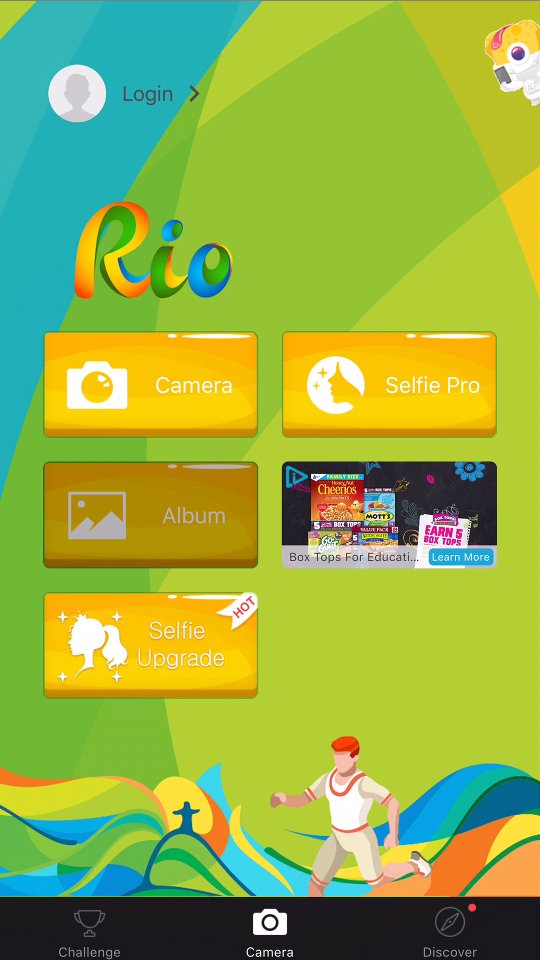}}
        \caption{Original screen-shot}
        \label{fig:original}
    \end{subfigure}%
	~
    \begin{subfigure}[t]{0.24\textwidth}
        \centering
        \frame{\includegraphics[width=\textwidth,clip=true,trim=0 0 5 0]{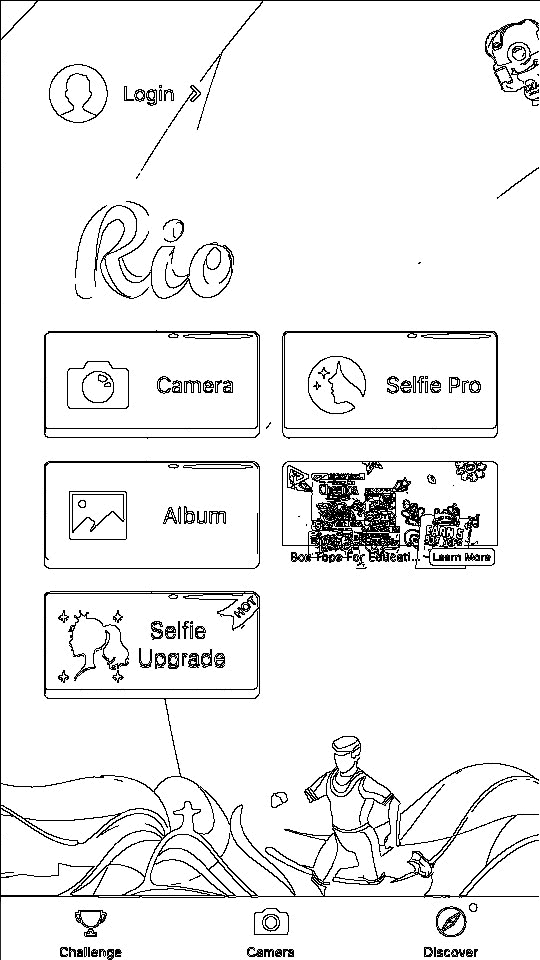}}
        \caption{Canny edges detection}
        \label{fig:canny}
    \end{subfigure}%
	~
	\begin{subfigure}[t]{0.24\textwidth}
        \centering
        \frame{\includegraphics[width=\textwidth,clip=true,trim=0 0 5 0]{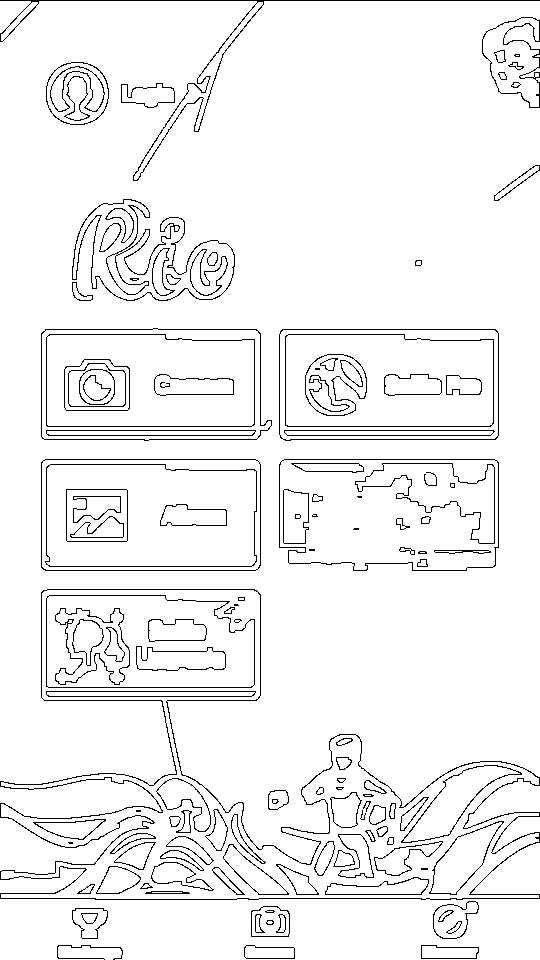}}
        \caption{Dilation and contours drawing}
        \label{fig:contour}
    \end{subfigure}%
	~
	\begin{subfigure}[t]{0.24\textwidth}
        \centering
        \frame{\includegraphics[width=\textwidth,clip=true,trim=0 0 5 0]{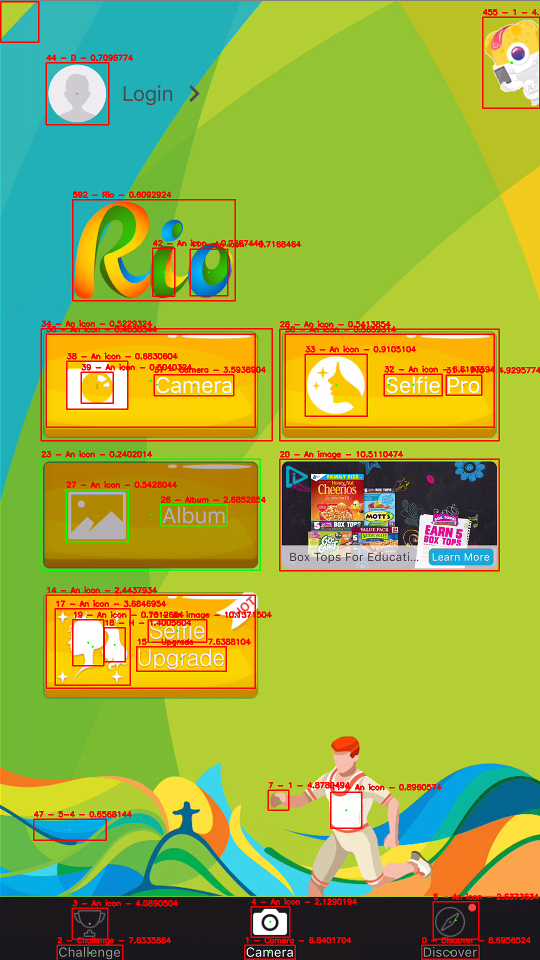}}
        \caption{GUI elements detection}
        \label{fig:GUIdectecttracking}
    \end{subfigure}%
	\caption{GUI element detection}
    \vspace{-4mm}
	\label{fig:detectFlow}
\end{figure*}

\section{Approach}
\label{sec:framework}

In this session, we propose {\tool} (\textbf{Min}ing V\textbf{i}sual L\textbf{o}g of Software Applicatio\textbf{n}), a framework to address the core problems of mining visual log of software. In general, {\tool} contains novel techniques and algorithms based on object recognition and tracking in computer vision/video processing to i) recognize GUI elements (e.g. buttons, sliders, or menu items) in the images capturing them, ii) infer the changes occurring on those GUI elements and then the user interactions causing such changes, and iii) generate the corresponding sequence of GUI interactions involving those elements and changes.

Figure~\ref{fig:framework} illustrates {\tool}. The core components of {\tool} includes a Visual Log Recorder, a Detector of GUI elements and changes, a Detector of GUI-based user interactions, and a Miner of GUI usage patterns. Each component will be discussed in more details in next sections.   

\subsection{Visual Log Recorder}

The Visual Log Recorder is used to capture the graphic user interface of a software system during its normal usages or its GUI testing usages. Its main output is sequences of images capturing the GUI screen (i.e. visual log data) during the usage session which can be stored as individual images or videos. If possible, it also records additional information of user interactions such as keyboard events, mouse or touch positions, mouse button clicks, etc. Such information would provide additional hints for the detection of GUI elements and their changes, and for the recognition of user interactions.

Although there are many screen capturing software products available in the market, a custom-built recorder for visual log is more useful because of the following reasons. First, we need to protect private data of users. The Visual Log Recorder in our framework will have techniques to recognize users' private data (photos, messages, passwords, etc.) and anonymize it, or replace it with common, non-private data. Second, while a typical screen capturing program can record the screen (including the GUI elements and data), it would store the recordings as video clips of a fixed frame rate (around 30 frames/second). However, most of the time the GUI is just waiting for user input and not having any significant changes (i.e. having no user interactions). Therefore, most of recorded frames will be redundant. In contrast, our custom-built recorder only captures the screen when major GUI events happen, including when users perform interactions with the GUI. Therefore, we will have fewer redundant frames. In addition, the custom-built recorder records extra information of GUI events (e.g. key pressing and mouse clicking) and associates them with the captured images. Finally, available screen capturing software stores recordings in lossy format to reduce storage cost, which introduces noise and thus, makes the mining process more difficult. In contrast, our recorder can store screen images in lossless format to avoid that issue.

\subsection{Detector of GUI elements}

As discussed in two motivating examples in Section I, the core problem of mining visual log of software is to \emph{recognize GUI elements and user interactions} with them. That is, given a visual log as a sequence of images capturing the graphical user interface of a software application during a usage session (e.g. a screen video clip), we need to infer what elements of the software's GUI appear in those images and what actions the users perform on those elements. From those, we can generate test scripts, mine frequent sub-sequences and other kinds of information.

{\tool} has a GUI Element Detector for ``detecting GUI elements and changes''. The input it receives from the Visual Log Recorder includes sequences of screen images. Its output is the GUI elements it recognized in each image (e.g. their types, positions, shape, and borders) and their changes (e.g. in color, position, and shape) in the image sequences. To recognize the GUI elements and changes, this component employs several algorithms designed based on computer vision (object recognition and tracking) and recognition rules manually designed specially for software GUI elements based on heuristics learned by studying those software GUI elements.

\subsection{Detector of user interactions with GUI elements}

This component infers the user interactions with GUI elements in a given visual log (i.e. a sequence of images associating with a sequence of GUI input events). It receives GUI input events provided by the Visual Log Recorder and GUI elements and changes detected by the GUI Element Detector. It also employs recognition rules that are designed manually and specially for software GUI elements and interactions to infer the user interactions. For example, if it recognizes the label of a button changing its color in the presenting of a user touch, it infers that the user has clicked on the button. The output of this component is a sequence of user interactions, i.e. each item describes a user action to a GUI element.

\subsection{Miner of GUI usage pattern}
When sequences of user interactions are extracted from a large collection of visual logs, we can learn usage patterns from those sequences. We plan to capture usage patterns in statistical generative models like n-gram, hidden Markov Model, or recurrent neural network~\cite{jurafsky_2009_speech}. Once trained, such models can estimate the occurring probability of any given sequence of user interaction, thus it can recognize usage patterns (i.e. sequences with extremely high probability of occurring) and usage anomalies (e.g. sequences with extremely low probability of occurring).

\section{Empirical study}

To design the algorithms for detecting GUI elements and user interactions, we performed a preliminary empirical study on the characteristics of GUI elements. We limit the scope of this study to photo apps for mobile phones because i) they would have the most intuitive and easy-to-use GUI and ii) their user base would be very large and highly diverse. We selected two popular photo apps \code{Camera 360} and \code{Photo Wonder} for this study.

\subsection{Data collection}

We designed 10 photo editing scenarios and performed them in those two apps on three different phone models (iPhone 6 Plus, 6, and 5s, each has a different screen size). While performing editing tasks, we recorded the screen of those apps. After that, we extracted screen shots (i.e. frames) from the recorded videos when major GUI events happened (e.g. clicking on a button or a list item) and selected a random sample of those screen-shots. Our final dataset contains 200 screen shots and 2,804 GUI elements.    

Then we manually inspected the selected sample. For each frame, we identified all visible GUI elements and collected their information. First, we classified them into three types: \code{Text} (e.g. captions of GUI elements, displayed strings), \code{Icon} (an image or drawing that can invoke a functionality), and \code{Comb} (combinations of text labels and icons, i.e. a compounded GUI elements). We also recorded their positions, bounds, shapes, and directions (e.g. vertically or horizontally). For example, Figure~\ref{fig:original} is the home screen of \code{Camera 360} running on an iPhone 6 Plus. We can see the three types of element in one single button. The button \code{Camera} is a \code{Comb} element which contains an \code{Icon} element (looks like a camera) and a \code{Text} element (i.e. the word \code{Camera}).

\subsection{Findings}

We performed several analysis on the collected data. Here are the key findings from those analysis.

\paragraph{Distribution of GUI elements' types} 48.3\% of the identified GUI elements are \code{Text}, 39.4\% are \code{Icon}, and 12.3\% are \code{Comb}. This finding suggests that half of GUI elements are simple text. Because current Optical Character Recognition (OCR) techniques can recognize textual items in images with very high accuracy~\cite{willis_2006_orc}, we could employ them to recognize text-based GUI elements.

\paragraph{Shapes and directions of icons} 41.8\% of \code{Icon} elements have irregular shapes. 21.7\% have rectangle, 9.6\% have circle, and 10.6\% have arrow shapes. 98.1\% of icons are placed horizontally.

\paragraph{Shapes and directions of compounded elements} Shapes of most of compounded elements are circle (46.7\%) or rectangle (41.3\%). 87\% of them are placed horizontally.

\paragraph{Relative size of GUI elements} We computed the relative size of a GUI element as the ratio between the area of its bounding box and the screen resolution.

We found that \code{Text} and \code{Icon} elements have stable relative size. For example, in iPhone 5s, most (75\% and more) of those elements have relative size from 0.1\% to 0.4\%. In addition, the larger the screens are the smaller their relative sizes are. For example, \code{Text} elements have median relative size of 0.2\% in iPhone 5s and 0.15\% in iPhone 6 Plus.

Relative size of \code{Comb} elements is also stable but often much bigger than that of \code{Text} and \code{Icon} elements. For example, in iPhone 5s, most (75\% and more) of \code{Comb} elements have relative size from 1\% to 5\%.

Our findings suggest several ideas to design the algorithm to detect GUI elements. First, we could employ OCR to detect \code{Text} elements. Then, for two other types, we could use size, shape, and direction to detect and distinguish compounded elements from \code{Icon} elements.

\section{Detection of GUI elements}

\begin{figure}[t!]
\begin{lstlisting}
function DetectGUIElement(Image image)
  Use OCR to detect text elements
  Apply edge detection (Canny algorithm)
  Apply dilation
  Detect contours in image
  Filter out unlikely contours
  Match contours as GUI elements 
\end{lstlisting}
\caption{Detection of GUI elements}

\vspace{-4mm}
\label{fig:elementDetection}
\end{figure}

Inspired by the work of Nguyen~\emph{et al.}~\cite{TuanASE15} and our empirical study, we propose an algorithm in Figure~\ref{fig:elementDetection} to detect GUI elements in a given screen-shot. The key idea of our algorithm is to use computer vision techniques (edges detection, dilation, and contours detection~\cite{gonzalez_2006_image}) to detect the contours in the image and then filter contours that are unlikely.    

Figure~\ref{fig:detectFlow} illustrates this algorithm. The first screen-shot shows the original homepage of \code{Camera 360}, the second shows the result of \code{Canny} edges detection, the third shows the result of contours extraction after dilatation was applied. The final image shows the detected GUI elements. 

As seen in Figure~\ref{fig:detectFlow}, some detected contours are not actually GUI elements. They might be distinctively visible as part of a larger image or simply artifacts created by the process of dilation and contours detection. Based on our empirical study, we propose the following rules to eliminate such negative candidates:

\begin{enumerate}  
\item If a contour is too small (e.g. relative size less than 0.01\%), it won't correspond to a GUI element. (GUI elements need to be large enough to be visible and touchable).

\item If a contour is too large (e.g. relative size more than 1\%) and does not have rectangle or circle shape or is not placed horizontal or vertical nicely, it is unlikely a \code{Comb} GUI element and should be ignored.
\end{enumerate}

\section{Work in process}

We are implementing our framework and improving the algorithm for GUI element detection. We expect that information of changes and motions could further enhance the detection. For example, GUI  elements moving together (e.g. icons in a list that could be swiped side way) are likely to belong to the same group, or elements having colors that changes together (e.g. an icon and the accompanied text) are likely representing the same function. 

We are also designing algorithms for detecting user interactions, which is also based on changes in detected GUI elements' properties (color, shape, position, size, etc.). For example, a change in color and/or shape of an element could mean a user click. However, there is a possibility that the change is not triggered by a user interaction but a routine automatic animation. To solve this, we could use the extra GUI input event information to infer the \emph{major events}. A major event contains enough changes in  objects' properties may indicate view or view fragment changes. If changes of a few related elements were detected before such a major event, we could infer that a user interaction has been performed. We could record such user interaction with text label, image from icon, or value changes (for example a user could move a slider that changes a certain parameter indicates by a changing number). From each visual log, we will be able to collect a sequence of user interaction by recording one detected single action at a time.



\bibliographystyle{abbrv}
\bibliography{visuallog}
\end{document}